\documentclass[aps,pra,twocolumn,amsmath,amssymb]{revtex4}%
\usepackage{graphics}
\usepackage{amsmath}
\usepackage{graphicx}
\usepackage{amsfonts}
\usepackage{amssymb}
\usepackage{float}
\usepackage{longtable}
\usepackage{epsfig}
\usepackage{latexsym}
\usepackage{theorem}
\usepackage{bbm}
\usepackage{bm}
\usepackage{verbatim}
\DeclareMathOperator\arcsinh{arcsinh}
\begin{document}
\title{Hacking Alice's box in continuous variable quantum key distribution}
\author{Jason Pereira, Stefano Pirandola}
\affiliation{Computer Science and York Centre for Quantum Technologies, University of York,
York YO10 5GH, UK}

\begin{abstract}
Security analyses of quantum cryptographic protocols typically rely on certain
conditions; one such condition is that the sender (Alice) and receiver (Bob)
have isolated devices inaccessible to third parties. If an eavesdropper (Eve)
has a side-channel into one of the devices, then the key rate may be sensibly
reduced. In this paper, we consider an attack on a coherent-state protocol,
where Eve not only taps the main communication channel but also hacks Alice's
device. This is done by introducing a Trojan horse mode with low mean number
of photons $\bar{n}$ which is then modulated in a similar way to the signal
state. First we show that this strategy\ can be reduced to an attack without
side-channels but with higher loss and noise in the main channel. Then we show
how the key rate rapidly deteriorates for increasing photons $\bar{n}$, being
halved at long distances each time $\bar{n}+1$ doubles. Our work suggests that
Alice's device should also be equipped with sensing systems that are able to
detect and estimate the total number of incoming and outgoing photons.

\end{abstract}
\maketitle

\section{Introduction}

Quantum information science~\cite{Watrous,Hayashi,HolevoBOOK} is
advancing at a rapid pace. The progress of quantum
computing~\cite{NC00} threatens to make current, classical
cryptography insecure. Quantum key distribution
(QKD)~\cite{rev1,rev2,rev3} is a possible solution to this
problem, offering provable information security based on physical
principles. It is possible to design QKD protocols that ensure
that any eavesdropper can hold only an arbitrarily small amount of
information about the message sent. This holds true regardless of
how advanced the eavesdropper's technology is.

Security proofs for QKD protocols have a few assumptions that must hold in
order for them to be valid \cite{pironio_device-independent_2009}. The two
trusted parties (Alice and Bob) must have isolated devices, which are
inaccessible to the eavesdropper (Eve). The devices should be fully
characterised, so that an adversary cannot exploit device imperfections to
acquire information about the key or to alter the trusted parties' estimations
of the quantum channel properties. The trusted parties must also have an
authenticated (but not secure) classical channel; an eavesdropper can listen
in to classical communications along this channel, but cannot alter them. If
we relax any of these conditions, the secure key rate for a protocol may change.

Current commercial implementations of discrete variable (DV) protocols, such
as BB84 \cite{BB84} with decoy states \cite{decoy,lo_decoy_2005}, have been
shown to be vulnerable to a variety of attacks that exploit device
imperfections, such as \textquotedblleft side-channels\textquotedblright\ that
leak information from the trusted parties' devices to Eve
\cite{scarani_black_2009}. These attacks include detector blinding attacks
\cite{lydersen_hacking_2010}, time-shift attacks \cite{zhao_quantum_2008} and
Trojan horse attacks \cite{jain_trojan-horse_2014}.

A variety of attacks on continuous variable (CV) protocols have been proposed. In experimental realisations of QKD, the local oscillator (which is used by Bob to carry out his measurements) is often sent down the quantum channel; this introduces a vulnerability that an eavesdropper can exploit. H\"{a}seler et al. \cite{haseler_testing_2008} demonstrated that Eve could disguise an intercept and resend attack by replacing the signal state and the local oscillator with squeezed states. The wavelength-dependence of beamsplitters in Bob's setup can be exploited to engineer his measurement outcomes \cite{huang_quantum_2013,ma_wavelength_2013}. Altering the shape of the local oscillator pulse can allow an eavesdropper to change Bob's estimation of the shot-noise \cite{jouguet_preventing_2013,huang_quantum_2014}. Saturation attacks \cite{qin_quantum_2016,qin_homodyne_2018}, which push Bob's detectors out of the linear mode of operation, have also been proposed.

One way of avoiding attacks that exploit device imperfections is to use
device-independent QKD \cite{Ekert,pironio_device-independent_2009}. This is a
family of protocols that do not require Alice's and Bob's devices to be
trusted. Such protocols are immune from many side-channel attacks, but have
significantly lower key rates than protocols that require trusted devices.
Measurement-device independent (MDI) QKD protocols have been formulated for
both the DV \cite{samMDI2012,lo_measurement-device-independent_2012} and the
CV \cite{pirandola_high-rate_2015} cases, and have much higher key rates than
fully device-independent protocols. MDI-QKD removes threats from the
detector's point of view, but still assumes that the state-preparation devices are
completely trusted. Therefore, MDI-QKD is also subject to the quantum hacking
described in this paper.

Here we consider a Trojan horse attack, where Eve sends extra photons into
Alice's device, in order to gain information about the states being sent
through the main quantum channel without disturbing the signal state. This type of attack was first considered in depth by Vakhitov et al.~\cite{vakhitov_large-pulse_2001}. Such an
attack may be used in DV protocols~\cite{vinay_burning_2018}, in order to
distinguish decoy states from signal states or to gain information about
Alice's basis choice.

Gisin et al.~\cite{gisin_trojan_2006} described how reflectometry could be used by Eve to gain information about Alice's phase modulator settings and analytically calculated the information leakage in terms of the photon number of the state received by Eve after the side-channel. They assumed attenuation of the side-channel mode by Alice and showed that the information leakage is reduced if Alice can randomise the phase of the side-channel mode.

Lucamarini et al.~\cite{lucamarini_practical-security_2015} calculated the secret key rate for BB84, with and without decoy states, in the presence of a Trojan horse side-channel, in terms of the photon number of the state received by Eve. They then bounded the incoming photon number in terms of the Laser Induced Damage Threshold (LIDT) of the optical fibre and the time for which Alice's device gate is open, assuming that the Trojan horse photons are sent in via the main channel, whilst the gate is open. Based on this constraint, they designed an architecture to passively limit the photon number of the received state, and hence the information leakage.

Tamaki et al.~\cite{tamaki_decoy-state_2016} found general analytical expressions for the information leakage of DV protocols due to Trojan horse attacks, in terms of the actions of the phase and intensity modulators. This allows the secret key rate of a general DV protocol in the presence of a Trojan horse side-channel to be calculated, as long as the phase and intensity modulators are well-characterised.

Here we assume a CV protocol based on the modulation of
coherent states~\cite{hetPROT}, so that the attack is against the modulator. The experimental viability of carrying out a Trojan horse attack on the commercial CV system SeQureNet has previously been considered~\cite{jain_risk-analysis_2014}.

More precisely we assume that Eve is both hacking Alice's device
with $\bar {n}$ mean photons per run and tapping the main quantum
channel between Alice and Bob, which can be assumed to be a
thermal-loss channel. This joint eavesdropping strategy can be
reduced to a side-channel-free attack but where the main quantum
channel has higher loss and noise. In this way we can compute the
secret key rate and how it varies in terms of the mean photons
$\bar{n}$. In particular, we show that, at long distances, the key
rate is halved each time $\bar{n}+1$ doubles. This means that
inserting just a few hacking photons into Alice's setup can
seriously endanger the security of the protocol. As a result of
our analysis, we conclude that the presence of these extra photons
should be actively monitored in any practical implementation of CV
QKD.

\section{Results}

\subsection{General scenario}

We consider two parties, Alice and Bob, who are trying to establish a secret
key, with a third party, Eve, trying to gain information about the secret key.
Alice initiates a coherent state protocol
\cite{hetPROT,weedbrook_gaussian_2012}. This involves her displacing a vacuum
state by a Gaussian-distributed random (two-dimensional) variable, $\alpha$. In real implementations, this displacement is generally carried out by independently modulating the phase and the intensity, so that the overall displacement has a Gaussian distribution.
She then sends the displaced vacuum state (called the signal state) to Bob,
via a quantum channel. Bob then carries out a heterodyne measurement on the
signal state, to obtain a value $\beta$. This process is repeated several
times. Alice and Bob compare some of their values via a classical
communication channel in order to establish the transmittance, $\eta$, and
excess noise, $\epsilon$, of the channel. Bob and Alice then establish a
secret key based on their shared knowledge of Bob's values (this is called
reverse reconciliation).

Whilst the signal states are in the main quantum channel, we allow Eve to
enact any unitary operation upon them. We assume that Eve can listen in on all
classical communication between Alice and Bob (but cannot alter it). She can
then store all states involved in the operation (except for the signal state)
in a quantum memory and carry out an optimal measurement on them after all
quantum and classical communication has been completed, in order to gain
information about Bob's values. Alice and Bob therefore assume that all of the
noise and loss of the channel has been caused by Eve's unitary operations and
try to bound the maximum knowledge that Eve could have obtained about Bob's
values. As long as Alice has more information about Bob's values than Eve, it
is possible for Alice and Bob to obtain a secret key.

If Eve is only able to access the main channel and is not able to access Alice
or Bob's devices in any way, the optimal attack on the signal state for a
given attenuation and noise is an entangling
cloner~\cite{grosshans_virtual_2003}. The secret key rate for this case has
been calculated~\cite{usenko_trusted_2016}. Here we instead consider the case
where Eve also has access to part of Alice's device via a side-channel. Eve
can send a Trojan horse mode into Alice's device, which will be displaced by
$\alpha$ in the same way as the signal state. This side-channel mode contains
an average number of photons $\bar{n}$, and we assume that Alice is able to
monitor these photons and estimate their number. This will not be the case for most current CV-QKD implementations, especially since certain potential Trojan horse side-channels may not have been identified yet, so additional quantum metrological tools must be placed inside Alice’s box in order for this assumption to be met. To represent Eve's Trojan
horse mode, we assume it is part of a two-mode squeezed vacuum (TMSV)
state~\cite{weedbrook_gaussian_2012} with squeezing $r$, so that $\bar
{n}=\sinh^{2}r$. This is an active attack when $\bar{n}>0$ and it is a passive
one when $\bar{n}=0$, meaning that we just have a leakage mode from Alice's device.

\begin{figure}[ptb]
\vspace{0.15cm} \centering
\includegraphics[width=1\linewidth]{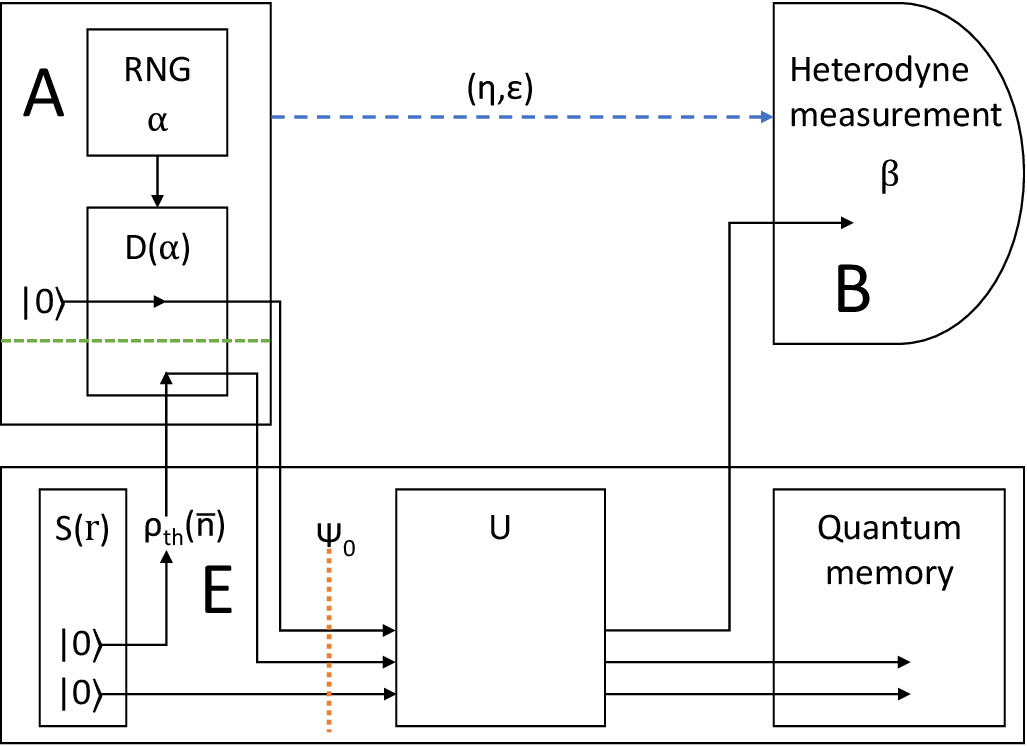}\caption{The channel setup under
consideration. A is Alice's device, B is Bob's device and E is Eve's device.
The dashed green line marks the part of Alice's device that is accessible to
Eve. Eve sends one mode of a TMSV state into Alice's device to be displaced by
$\alpha$ in the same way as the signal state. Alice knows the average photon
number, $\bar{n}$, of Eve's state. The (displaced) squeezed vacuum modes and
the signal state form the state $\psi_{0}$. Eve enacts a unitary on this total
state and any ancillary modes, then sends the signal state to Bob and stores
the remaining modes in a quantum memory. Bob carries out a heterodyne
measurement on the signal state, obtaining $\beta$. We find the key rate
assuming that the main channel is a thermal channel, with transmittance $\eta$
and excess noise $\epsilon$, as represented by the blue dashed arrow.}%
\label{fig:init_setup}%
\end{figure}

Recently, a side-channel on CV-QKD based on leakage from a multimode modulator
was considered by Derkach et al.~\cite{derkach_continuous-variable_2017}, building on their previous work~\cite{derkach_preventing_2016}. These works considered leakage modes prior to and after modulation of the signal state, for both the coherent state and the squeezed state protocols.
However, these authors did not consider side-channels that allow Eve to send photons into Alice's device (non-zero values of $\bar{n}$). They also considered homodyne, rather than
heterodyne, measurements by the receiver. In this paper, we will consider a
more general scenario, where the hacking of Alice's device is active,
therefore involving the use of two-mode squeezing, so that $\bar{n}>0$ photons enter
the device. We analyse the security when the side-channel mode is modulated by
$\alpha$, exactly as the signal mode is (we later generalise to the case where
its modulation is $m\alpha$). See Fig.~\ref{fig:init_setup} for an overview of
the situation.

To find the secret key rate in reverse reconciliation, we need to calculate
the mutual information between Alice and Bob $I(\alpha:\beta)$ and that
between Eve and Bob. The latter is upper-bounded by the Holevo bound
$I(E:\beta)$, which can be calculated as the reduction in entropy of Eve's
output state when conditioned by Bob's value, $\beta$. We upper-bound Eve's
knowledge of Bob's state by assuming that all noise and loss experienced by
the signal state is due to Eve enacting unitary operations on the signal state
and some ancillary modes, which are then stored in a quantum memory.

\subsection{Reduction of the attack}

If there are no side-channels, Eve's Holevo bound can be calculated by
assuming that the signal state is entangled with some state held by Alice, and
that $\alpha$ is the result of a heterodyne measurement on a TMSV
state~\cite{grosshans_virtual_2003}. In the presence of our side-channel, the
initial state held by Eve prior to her enacting the main channel is tripartite
and composed of the signal mode and Eve's side-channel modes. Our first step
must be to determine the first and second moments of this state $\psi_{0}$
(see Fig.~\ref{fig:init_setup}). We label the initial first moment vector
$X_{0}$ and the initial second moment (covariance) matrix $V_{0}$. For a fixed
value of $\alpha$, we have the conditional state $\psi_{0}|\alpha$ which is
the tensor product of a coherent state $\left\vert \alpha\right\rangle
\left\langle \alpha\right\vert $ and a TMSV state where one of the modes has
also been displaced by $\alpha$. The conditional moments are given by
\begin{equation}
X_{0}|\alpha=%
\begin{pmatrix}
\alpha\\
\alpha\\
0
\end{pmatrix}
,~~V_{0}|\alpha=%
\begin{pmatrix}
\mathbf{1} & \mathbf{0} & \mathbf{0}\\
\mathbf{0} & \cosh{2r}\mathbf{1} & \sinh{2r}\mathbb{Z}\\
\mathbf{0} & \sinh{2r}\mathbb{Z} & \cosh{2r}\mathbf{1}%
\end{pmatrix}
,
\end{equation}
where $\mathbf{1}$ is the one-mode identity matrix, $\mathbf{0}$ is the
one-mode zero-matrix, and $\mathbb{Z}$ is the Pauli Z-matrix.

In order to find the elements of $V_{0}$, we add the expectation value of
$X_{0}|\alpha\cdot X_{0}|\alpha^{T}$ to $V_{0}|\alpha$. Using $\left\langle
\alpha\right\rangle =0$ and $\left\langle \alpha^{2}\right\rangle =\mu$, we
find
\begin{equation}
X_{0}=%
\begin{pmatrix}
0\\
0\\
0
\end{pmatrix}
,~V_{0}=%
\begin{pmatrix}
(\mu+1)\mathbf{1} & \mu\mathbf{1} & \mathbf{0}\\
\mu\mathbf{1} & (\mu+\cosh{2r})\mathbf{1} & \sinh{2r}\mathbb{Z}\\
\mathbf{0} & \sinh{2r}\mathbb{Z} & \cosh{2r}\mathbf{1}%
\end{pmatrix}
.
\end{equation}
From the covariance matrix $V_{0}$ we can compute the three symplectic
eigenvalues~\cite{weedbrook_gaussian_2012}
\begin{align}
v_{1}  &  =1,\\
v_{2}  &  =\mu+\sqrt{1+\mu+\mu^{2}+\mu\cosh{2r}},\\
v_{3}  &  =-\mu+\sqrt{1+\mu+\mu^{2}+\mu\cosh{2r}},
\end{align}
and compute the entropy of the total state as~\cite{weedbrook_gaussian_2012}
$S(\psi_{0})=\sum\nolimits_{k=1}^{3}g(v_{k})$
where~\cite{grosshans_collective_2005}
\begin{align}
g(x)  &  =\frac{x+1}{2}\log_{2}\frac{x+1}{2}-\frac{x-1}{2}\log_{2}\frac
{x-1}{2}\\
&  \overset{x\gg1}{\rightarrow}\log_{2}\frac{ex}{2}+O(x^{-1}). \label{asym}%
\end{align}

\begin{figure}[ptb]
\vspace{0.15cm} \centering
\includegraphics[width=1\linewidth]{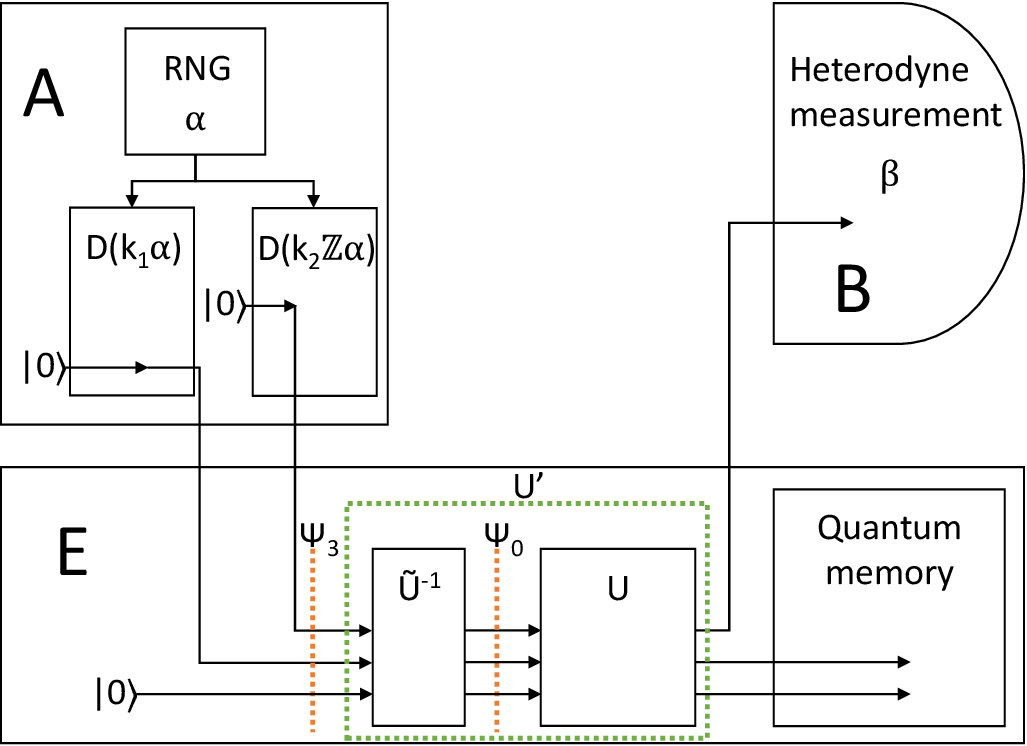}\caption{An equivalent channel to
the setup in Fig.~\ref{fig:init_setup}. Alice draws a two-dimensional
variable, $\alpha$, from a Gaussian distribution then displaces one vacuum
state by $k_{1}\alpha$ and another by $k_{2}\mathbb{Z}\alpha$. The first mode
is sent through the main channel to Bob as the signal state and the second
mode is leaked to Eve. The equivalence can be seen from the fact that Eve can
get the initial state from Fig.~\ref{fig:init_setup}, $\psi_{0}$, by enacting
the unitary $\tilde{U}^{-1}$, and can then enact the same arbitrary unitary,
$U$. We can regard this as Eve enacting a single combined unitary, $U^{\prime
}$.}%
\label{fig:equiv1}%
\end{figure}

\begin{figure}[ptb]
\vspace{0.1cm} \centering
\includegraphics[width=1\linewidth]{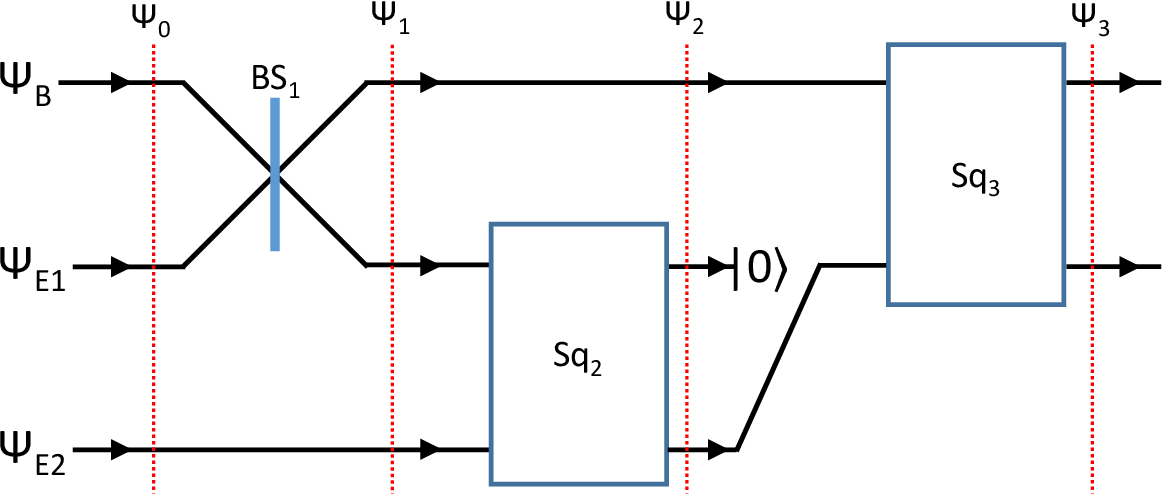}\caption{A circuit that converts
the initial (pre-main channel) state from the setup in
Fig.~\ref{fig:init_setup} into the initial state from the setup in
Fig.~\ref{fig:equiv1}. This shows that the two channel setups have the same
key rate, since Eve can enact any unitary operation and hence is able to
convert one into the other. We label this entire circuit $\tilde{U}$. Eve can
also enact the inverse, $\tilde{U}^{-1}$. $\psi_{B}$ denotes the signal state,
$\psi_{E1}$ denotes Eve's squeezed state that enters the side-channel and
$\psi_{E2}$ denotes Eve's idler state. $\mathrm{BS}_{1}$ is a balanced
beamsplitter and $\mathrm{Sq}_{2}$ and $\mathrm{Sq}_{3}$ are two-mode
squeezers. $\mathrm{BS}_{1}$ moves all of the displacement onto the first
mode, such that Eve's states are no longer displaced, $\mathrm{Sq}_{2}$
unsqueezes Eve's states such that one of the modes becomes a pure vacuum state
and $\mathrm{Sq}_{3}$ unsqueezes the signal state and Eve's remaining mode
such that they become pure displaced vacuum states.}%
\label{fig:circuit}%
\end{figure}

The fact that $v_{1}=1$ tells us that there is a symplectic transformation
that reduces $\psi_{0}$ to a tensor product of a two-mode state and a vacuum
state. We can build on this observation and reduce the number of modes. In
fact, we may show the reduction to the setup in Fig.~\ref{fig:equiv1}, which
only involves the signal mode, modulated by $k_{1}\alpha$ (with $k_{1}%
>1$),\ and a single Trojan horse mode, modulated by $k_{2}\mathbb{Z}\alpha$
(with $k_{2}$ real). We can design a Gaussian unitary $\tilde{U}$
that converts the initial state $\psi_{0}$ from Fig.~\ref{fig:init_setup} into
the initial state $\psi_{3}$ from Fig.~\ref{fig:equiv1}. This unitary
operation $\tilde{U}$ is the optical circuit shown in Fig.~\ref{fig:circuit},
where we have labelled the signal state as $\psi_{B}$, Eve's squeezed state
that enters the side-channel as $\psi_{E1}$ and Eve's idler state (the
squeezed state that does not enter the side-channel) as $\psi_{E2}$.

To see how the circuit transforms the state, we examine it after each of the
three optical components; we label the states after each component with the
subscripts 1, 2 and 3. $\psi_{i}$ has first moments vector $X_{i}$ and
covariance matrix $V_{i}$. The conditional state $\psi_{i}|\alpha$\ is
associated to $X_{i}|\alpha$ and $V_{i}|\alpha$.\ The symplectic matrix of the
i\textsuperscript{th} component is $S_{i}$ and it characterises the
transformation of the state from $\psi_{i-1}$ to $\psi_{i}$ as follows:
$V_{i}=S_{i}V_{i-1}S_{i}^{T}$ and $X_{i}=S_{i}X_{i-1}$.

The first component is a balanced beamsplitter, acting on the signal state and
Eve's side-channel mode. This sets the quadratures for Eve's side-channel mode
to 0. It has symplectic matrix
\begin{equation}
S_{1}=%
\begin{pmatrix}
\frac{1}{\sqrt{2}}\mathbf{1} & \frac{1}{\sqrt{2}}\mathbf{1} & \mathbf{0}\\
-\frac{1}{\sqrt{2}}\mathbf{1} & \frac{1}{\sqrt{2}}\mathbf{1} & \mathbf{0}\\
\mathbf{0} & \mathbf{0} & \mathbf{1}%
\end{pmatrix}
,
\end{equation}
and it results in the following moments for $\psi_{1}|\alpha$ and $\psi_{1}$
\begin{align}
X_{1}|\alpha &  =%
\begin{pmatrix}
\sqrt{2}\alpha\\
0\\
0
\end{pmatrix}
,\\
V_{1}|\alpha &  =%
\begin{pmatrix}
\cosh^{2}r\mathbf{1} & \sinh^{2}r\mathbf{1} & \frac{\sinh{2r}}{\sqrt{2}%
}\mathbb{Z}\\
\sinh^{2}r\mathbf{1} & \cosh^{2}r\mathbf{1} & \frac{\sinh{2r}}{\sqrt{2}%
}\mathbb{Z}\\
\frac{\sinh{2r}}{\sqrt{2}}\mathbb{Z} & \frac{\sinh{2r}}{\sqrt{2}}\mathbb{Z} &
\cosh{2r}\mathbf{1}%
\end{pmatrix}
,\\
V_{1} &  =V_{1}|\alpha\oplus2\mu%
\begin{pmatrix}
\mathbf{1} &  & \\
& \mathbf{0} & \\
&  & \mathbf{0}%
\end{pmatrix}
.
\end{align}

The second component is a two-mode squeezer, operating on Eve's modes such
that one of them becomes a vacuum state. Its squeezing parameter is given by
$r_{2}=\log\left(  \frac{\sqrt{2}\cosh{r}-\sinh{r}}{\sqrt{\cosh^{2}r+1}%
}\right)  $, and it has symplectic matrix
\begin{equation}
S_{2}=%
\begin{pmatrix}
\mathbf{1} & \mathbf{0} & \mathbf{0}\\
\mathbf{0} & \frac{\sqrt{2}\cosh{r}}{\sqrt{\cosh^{2}r+1}}\mathbf{1} &
-\frac{\sinh{r}}{\sqrt{\cosh^{2}r+1}}\mathbb{Z}\\
\mathbf{0} & -\frac{\sinh{r}}{\sqrt{\cosh^{2}r+1}}\mathbb{Z} & \frac{\sqrt
{2}\cosh{r}}{\sqrt{\cosh^{2}r+1}}\mathbf{1}%
\end{pmatrix}
.
\end{equation}
The moments of $\psi_{2}|\alpha$ and $\psi_{2}$ are given by%
\begin{align}
X_{2}|\alpha &  =%
\begin{pmatrix}
\sqrt{2}\alpha\\
0\\
0
\end{pmatrix}
,\\
V_{2}|\alpha &  =%
\begin{pmatrix}
\cosh^{2}r\mathbf{1} & \mathbf{0} & \sqrt{\cosh^{4}r-1}\mathbb{Z}\\
\mathbf{0} & \mathbf{1} & \mathbf{0}\\
\sqrt{\cosh^{4}r-1}\mathbb{Z} & \mathbf{0} & \cosh^{2}r\mathbf{1}%
\end{pmatrix}
,\\
V_{2} &  =V_{2}|\alpha\oplus2\mu%
\begin{pmatrix}
\mathbf{1} &  & \\
& \mathbf{0} & \\
&  & \mathbf{0}%
\end{pmatrix}
.
\end{align}

Note that one of the modes has become a vacuum state. Henceforth, we neglect
this mode and implicitly enact the identity operation on it. We now see that,
for fixed $\alpha$, the system is a displaced TMSV state. The third component
undoes the squeezing, leaving us with two displaced vacuum states. Its
squeezing parameter is given by $r_{3}=-\arcsinh\left(  \frac{\sinh{r}}%
{\sqrt{2}}\right)  $ and it has symplectic matrix
\begin{equation}
S_{3}=%
\begin{pmatrix}
\frac{\sqrt{\cosh^{2}{r}+1}}{\sqrt{2}}\mathbf{1} & -\frac{\sinh{r}}{\sqrt{2}%
}\mathbb{Z}\\
-\frac{\sinh{r}}{\sqrt{2}}\mathbb{Z} & \frac{\sqrt{\cosh^{2}{r}+1}}{\sqrt{2}%
}\mathbf{1}%
\end{pmatrix}
.
\end{equation}
The moments of $\psi_{3}|\alpha$ and $\psi_{3}$ are%
\begin{equation}
X_{3}|\alpha=%
\begin{pmatrix}
k_{1}\alpha\\
k_{2}\mathbb{Z}\alpha
\end{pmatrix}
,~V_{3}|\alpha=%
\begin{pmatrix}
\mathbf{1} & \mathbf{0}\\
\mathbf{0} & \mathbf{1}%
\end{pmatrix}
,
\end{equation}%
\begin{equation}
V_{3}=%
\begin{pmatrix}
(1+\mu(\cosh^{2}r+1))\mathbf{1} & -\mu\sqrt{\cosh^{4}r-1}\mathbb{Z}\\
-\mu\sqrt{\cosh^{4}r-1}\mathbb{Z} & (1+\mu\sinh^{2}r)\mathbf{1}%
\end{pmatrix}
,
\end{equation}
where we have set
\begin{equation}
k_{1}:=\sqrt{\cosh^{2}{r}+1},~k_{2}:=-\sinh{r}.
\end{equation}
This concludes the proof of equivalence between the setups in
Fig.~\ref{fig:init_setup} and Fig.~\ref{fig:equiv1}.

\begin{figure}[ptb]
\centering
\includegraphics[width=1\linewidth]{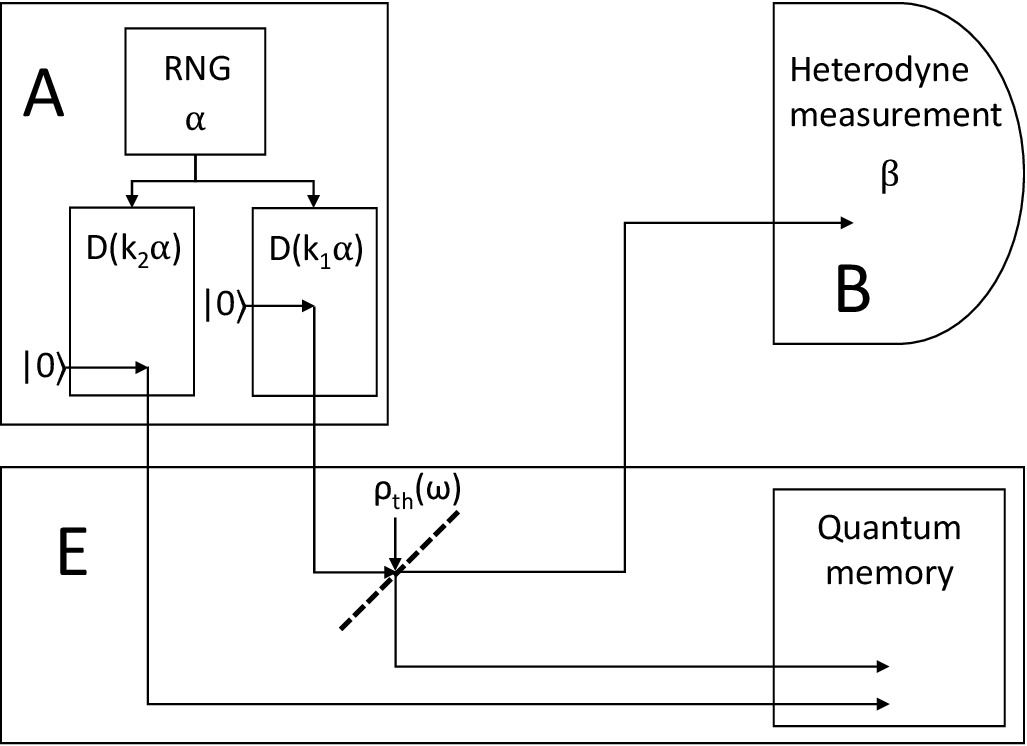}\caption{An alternative channel
setup that must give the same secret key rate as the setup in
Fig.~\ref{fig:equiv1} assuming the presence of a thermal-loss channel. The
difference between the two setups is that in Fig.~\ref{fig:equiv1}, the
x-quadrature of Eve's side-channel state is modulated by $k_{2}\alpha_{x}$ and
the p-quadrature is modulated by $-k_{2}\alpha_{p}$; in this figure,the
x-quadrature is still modulated by $k_{2}\alpha_{x}$ but the p-quadrature is
modulated by $k_{2}\alpha_{p}$. Since the two quadratures encode independent
variables and since the x-quadrature is not affected by the change, the mutual
informations arising from the measurement of the x-quadrature, $I_{AB}^{x}$
and $I_{EB}^{x}$, must be the same in each setup and hence the key rates must
be the same. We assume that Eve beamsplits the signal state with some thermal
state with variance $\omega$. This specific representation of Eve's unitary is
unique up to isometries on her output ancillas. In other words, if we fix the
channel to be thermal-loss, then its dilation into a beams-splitter with an
environmental thermal state is fixed up to unitaries acting over Eve's entire
output Hilbert space~\cite{pirandola_characterization_2008}.}%
\label{fig:equiv2}%
\end{figure}

We note that the two components (quadratures) of $\alpha$ are uncorrelated
with each other and have the same variance. Let us also assume that the two
quadratures of Bob's outcome ($\beta$) are also uncorrelated with each other
and have the same variance. This is certainly the case in the presence of a
thermal-loss channel, characterised by a transmittance $\eta$ and an excess
noise $\epsilon$, which is the most typical scenario in QKD. Next, we show
that the setup in Fig.~\ref{fig:equiv1} has the same key rate as the setup in
Fig.~\ref{fig:equiv2}, in which the signal mode is modulated by $k_{1}\alpha$
and the side-channel mode is modulated by $k_{2}\alpha$ (rather than by
$k_{2}\mathbb{Z}\alpha$). Note that in Fig.~\ref{fig:equiv2}, we have also
imposed that the general unitary results in a thermal-loss channel.

Since we assume that the main channel does not mix the quadratures, we can
treat the two quadratures of $\alpha$, which we denote as $\alpha_{x}$ and
$\alpha_{p}$, as independent variables that have been sent through the channel
and measured to give the independent variables $\beta_{x}$ and $\beta_{p}$
respectively. Let $I_{AB}^{x}$ ($I_{AB}^{p}$) denote the mutual information
between Alice and Bob arising from the measurement of the x-quadrature
(p-quadrature) and let $I_{EB}^{x}$ ($I_{EB}^{p}$) denote the maximum mutual
information between Eve and Bob arising from the measurement of the
x-quadrature (p-quadrature). Since the x and p quadratures of $\alpha$ and
$\beta$ are independent and identically distributed, $I_{AB}$ and $I_{EB}$ are
double $I_{AB}^{x}$ and $I_{EB}^{x}$ respectively.

Let $I_{AB}^{\prime}$, $I_{AB}^{\prime x}$, $I_{EB}^{\prime}$ and
$I_{EB}^{\prime x}$ be the counterparts of $I_{AB}$, $I_{AB}^{x}$, $I_{EB}$
and $I_{EB}^{x}$ respectively for the setup in Fig.~\ref{fig:equiv2}. It is
again true that $I_{AB}^{\prime}$ and $I_{EB}^{\prime}$ are double
$I_{AB}^{\prime x}$ and $I_{EB}^{\prime x}$ respectively. Further, since the
quadratures are independent and the x-quadratures of Eve's states are not
affected by the change in setup (the only difference is that the p-quadrature
of Eve's side-channel mode is modulated by $k_{2}\alpha_{p}$ rather than by
$-k_{2}\alpha_{p}$), $I_{AB}^{x}$ must be the same as $I_{AB}^{\prime x}$.
This means that $I_{AB}$ is the same as $I_{AB}^{\prime}$ and $I_{EB}$ is the
same as $I_{EB}^{\prime}$. Note that this holds for all channels (not just
thermal channels) that do not mix the quadratures and so the $\mathbb{Z}$
matrix in Fig.~\ref{fig:equiv1} can be neglected for any such channel.

\begin{figure}[ptb]
\vspace{0.15cm} \centering
\includegraphics[width=1\linewidth]{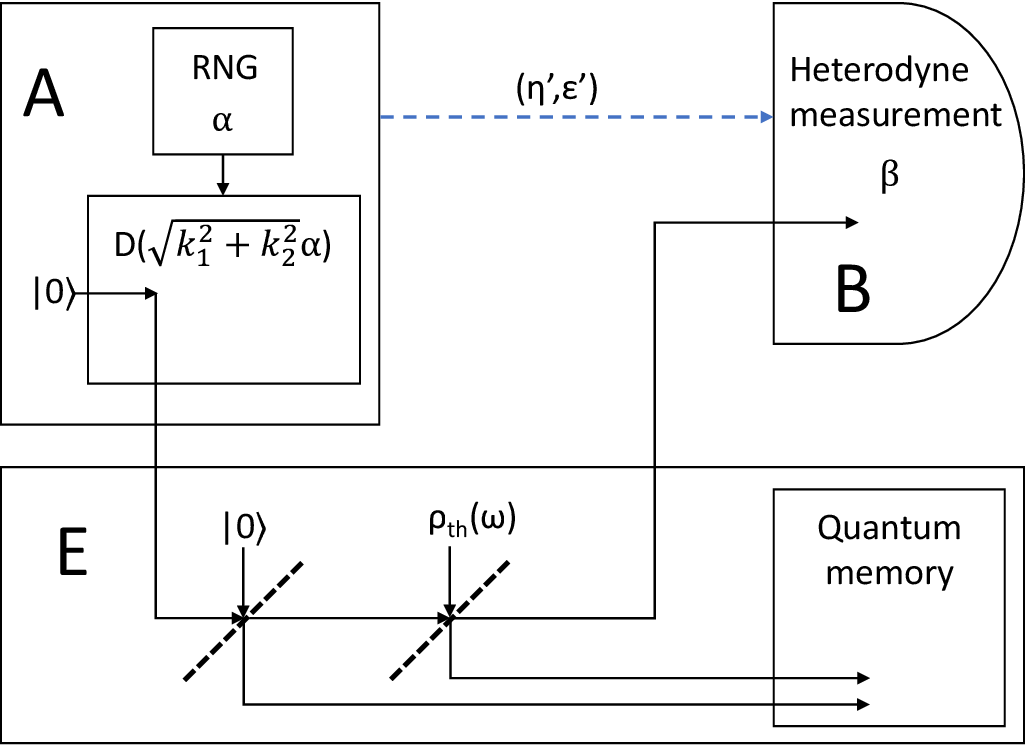}\caption{This is a setup without
a side-channel that must give the same secret key rate as the setup with the
side-channel. The variance of Alice's variable in this setup is higher than
the actual variance of $\alpha$, and the channel transmittance for this setup
is lower than the observed channel transmittance, $\eta$. The channel for this
setup can be regarded as a thermal channel with parameters $\eta^{\prime}$ and
$\epsilon^{\prime}$ (represented by the blue, dashed arrow).}%
\label{fig:equiv3}%
\end{figure}

Hence, the setup in Fig.~\ref{fig:equiv2} must give the same key rate as the
setup in Fig.~\ref{fig:equiv1}, and therefore the setup in
Fig.~\ref{fig:init_setup}. The setup in Fig.~\ref{fig:equiv2} is equivalent to
a main channel setup with a higher initial modulation and a lower effective
transmittance. The equivalent main channel attack is shown in
Fig.~\ref{fig:equiv3}. The signal state is modulated by $k\alpha$, where
\begin{equation}
k=\sqrt{k_{1}^{2}+k_{2}^{2}}=\sqrt{2}\cosh r=\sqrt{2(\bar{n}+1)},
\end{equation}
and hence the modulation amplitude is $k^2\mu$. $k$ is a function of $\bar{n}$, which characterises the side-channel.
We note that $k_{1}$ and $k_{2}$ are functions only of $\bar{n}$. By choosing
an appropriate parameter for the beamsplitter in Fig.~\ref{fig:equiv3}, Eve
can get the initial state of Fig.~\ref{fig:equiv2}. We then effect a thermal
channel by beamsplitting with the thermal state with parameter $\omega$. We
can reduce both operations to a single beamsplitter operation with some other
thermal state $\omega^{\prime}$ (see Fig.~\ref{fig:equiv4}).

This allows us to calculate the key rate in the same way as a main channel
attack but with a higher \textquotedblleft effective modulation amplitude",
$\mu^{\prime}$, and a lower \textquotedblleft effective transmittance",
$\eta^{\prime}$. These effective parameters (the channel parameters that the trusted parties would calculate for the setup in Fig.~\ref{fig:equiv3}) are related to the measured values
of $\mu$ and $\eta$ by
\begin{equation}
\mu^{\prime}=k^{2}\mu,~~\eta^{\prime}=\frac{\eta}{k^{2}}.\label{k1}%
\end{equation}
The effective transmittance accounts for both beamsplitters and is the
transmittance that we would observe if, instead of a setup with a signal state
modulated by $\mu$ and a side-channel (as seen in Fig.~\ref{fig:init_setup}), we had a setup with a signal state
modulated by $\mu^{\prime}$ and no side-channel, with the same measured values
of $\beta$ (as seen in Fig.~\ref{fig:equiv3}). This was found by multiplying the transmissions of the two beamsplitters in Fig.~\ref{fig:equiv3}.

It is helpful to clarify the definition of the excess noise, $\epsilon$. To do
so, we introduce the random variable $n$: this is the total relative input
noise of $\beta$ around $\alpha$, including the vacuum noise. We can describe
$\beta$ in terms of $n$ as $\beta=\sqrt{\eta}(\alpha+n)$. Here $n$ is characterised
by its second moment $\left\langle n^{2}\right\rangle =1+(1-\eta
)/\eta+\epsilon$. We now find the effective excess noise, $\epsilon^{\prime}$ (as would be observed for the setup in Fig.~\ref{fig:equiv3}),
using the fact that we have the same measured $\beta$ values in all
representations. $\beta$ can be expressed in terms of effective parameters as
$\beta=\sqrt{\eta^{\prime}}(k\alpha+n^{\prime})$, where the second moment of
$n^{\prime}$ is now given by $\left\langle n^{\prime2}\right\rangle
=1+(1-\eta^{\prime})/\eta^{\prime}+\epsilon^{\prime}$. We then substitute in the definition of $\eta'$ and compare the
expressions for $\beta$, and so solve for $\epsilon^{\prime}$, i.e.,
\begin{equation}
\epsilon^{\prime}=\frac{\eta}{\eta^{\prime}}\epsilon=k^{2}\epsilon. \label{k2}%
\end{equation}

\subsection{Computation of the key rate}

\begin{figure}[ptb]
\vspace{0.15cm} \centering
\includegraphics[width=1\linewidth]{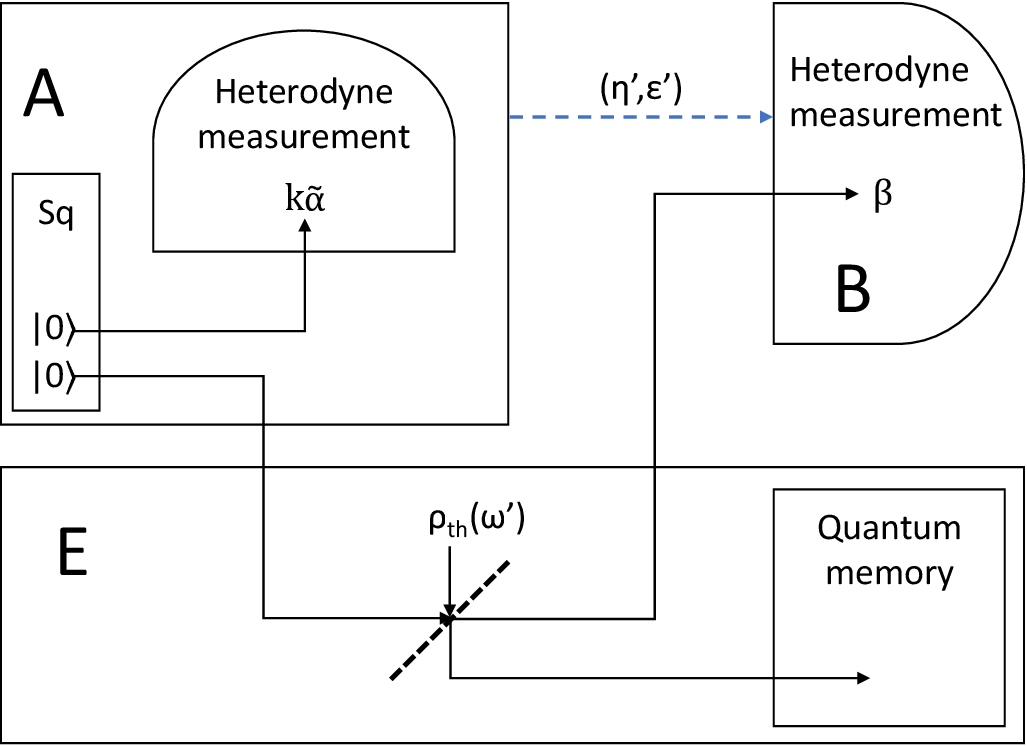}\caption{This is the
entanglement-based representation of the attack in Fig.~\ref{fig:equiv3}.
Alice heterodynes one half of a TMSV state to get the value $k\tilde{\alpha}$,
which linearly corresponds to $k\alpha$ (the displacement of the signal
state). The signal state enters the channel and is subject to some thermal
noise due to beamsplitting with one mode of an entangling cloner (the thermal
state $\omega^{\prime}$). It is then heterodyned by Bob, to obtain $\beta$.
The resultant state of Alice, Bob and Eve is pure. The channel between Alice
and Bob is a thermal channel, characterised by $\eta^{\prime}$ and
$\epsilon^{\prime}$; this is represented by the blue, dashed arrow.}%
\label{fig:equiv4}%
\end{figure}

To calculate the secret key rate for a main channel attack with a modulation
amplitude of $\mu^{\prime}$, a transmittance of $\eta^{\prime}$ and an excess
noise of $\epsilon^{\prime}$, we can use an entanglement-based representation
(rather than a prepare and measure representation)
\cite{grosshans_virtual_2003}. This representation is shown in
Fig.~\ref{fig:equiv4}, and is valid as long as $\mu>0$.

Alice heterodynes one mode of a TMSV state, obtaining the value $k\tilde
{\alpha}$ (and hence also the value of $\alpha$) and preparing the state
$\rho(k\alpha)$. She then sends the prepared signal state through the channel
to Bob, who heterodynes it to obtain $\beta$. In the channel, the signal state
is beamsplit with the thermal state $\rho_{th}(\omega)$. The total state
shared by Alice, Bob and Eve, which we denote $\rho_{ABE}$, is pure since Eve
holds the purification of the channel. This means that the entropy of Eve's
state, $\rho_{E}$, is equal to the entropy of the combined state of Alice and
Bob, $\rho_{AB}$. The combined state of Alice and Eve conditioned by some
value of $\beta$, $\rho_{AE}|\beta$, is also pure, so the entropy of Eve's
state conditioned by $\beta$, $\rho_{E}|\beta$, is equal to the entropy of
Alice's state conditioned by $\beta$, $\rho_{A}|\beta$.

The covariance matrix of $\rho_{AB}$ is
\begin{equation}
V_{AB}=%
\begin{pmatrix}
(\mu^{\prime}+1)\mathbf{1} & \sqrt{\eta^{\prime}\mu^{\prime}(\mu^{\prime}%
+2)}\mathbb{Z}\\
\sqrt{\eta^{\prime}\mu^{\prime}(\mu^{\prime}+2)}\mathbb{Z} & (\eta^{\prime
}(\mu^{\prime}+\epsilon^{\prime})+1)\mathbf{1}%
\end{pmatrix}
,
\end{equation}
the covariance matrices of the conditional states $\rho_{A}|\beta$ and
$\rho_{B}|\alpha$ are given by%
\begin{align}
V_{A}|\beta &  =\left(  \mu^{\prime}+1-\frac{\eta^{\prime}\mu^{\prime}%
(\mu^{\prime}+2)}{\eta^{\prime}(\mu^{\prime}+\epsilon^{\prime})+2}\right)
\mathbf{1,}\\
V_{B}|\alpha &  =(\eta^{\prime}\epsilon^{\prime}+1)\mathbf{1}.
\end{align}
We can calculate the symplectic eigenvalues of $V_{AB}$ using the formula
in~\cite{weedbrook_gaussian_2012}. The expressions for these eigenvalues can
be simplified by taking the asymptotic limit in $\mu$ (the limit as
$\mu\rightarrow\infty$). In this limit, $\mu^{\prime}\rightarrow\infty$ and
all other parameters stay the same. We assume that $\eta^{\prime}\leq1$, since
realistically, Eve will not enact a main channel that causes gain rather than
loss. We denote the two symplectic eigenvalues of $V_{AB}$ in this limit as
$v_{AB,1}^{\infty}$ and $v_{AB,2}^{\infty}$ and denote the symplectic
eigenvalue of $V_{A}|\beta$ in this limit as $v_{A|\beta}^{\infty}$. We find
these to be:
\begin{align}
v_{AB,1}^{\infty}  &  =1+\frac{\epsilon^{\prime}\eta^{\prime}}{1-\eta^{\prime
}},\\
v_{AB,2}^{\infty}  &  =\mu^{\prime}(1-\eta^{\prime}),\\
v_{A|\beta}^{\infty}  &  =\frac{2}{\eta^{\prime}}+\epsilon^{\prime}-1.
\end{align}

\begin{figure}[ptb]
\centering
\includegraphics[width=1\linewidth]{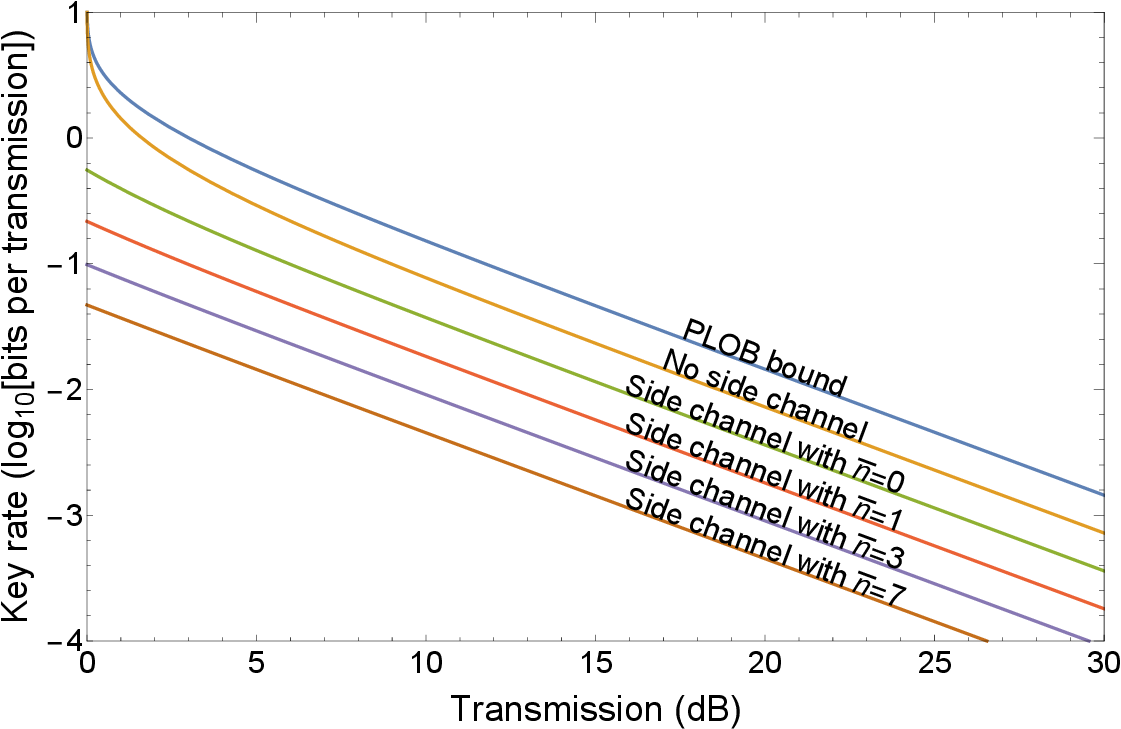}\caption{Plots of the secret
key rate (in logarithmic scale) versus channel transmission $\eta$ of the main
quantum channel, in the absence of excess noise (lossy channel rate). The top curve is the PLOB
bound \cite{PLOB}, which is the secret key capacity of the lossy channel,
i.e., the maximum key rate achievable over this channel by any point-to-point
QKD protocol in the absence of side-channels~\cite{TQC}. We then show the
ideal rate of the coherent state protocol~\cite{hetPROT} with no side
channels. Lower curves refer to the coherent state protocol in the presence of
a side-channel with an increasing number of photons $\bar{n}$, ranging from
the leakage mode case ($\bar{n}=0$) to more active hacking ($\bar{n}=1$, $3$,
$7$). As we can see, the key rate is always positive (for any value of
$\bar{n}$), but it quickly declines as $\bar{n}$ increases.}%
\label{fig:lossy}%
\end{figure}

\begin{figure}[ptb]
\centering
\includegraphics[width=1\linewidth]{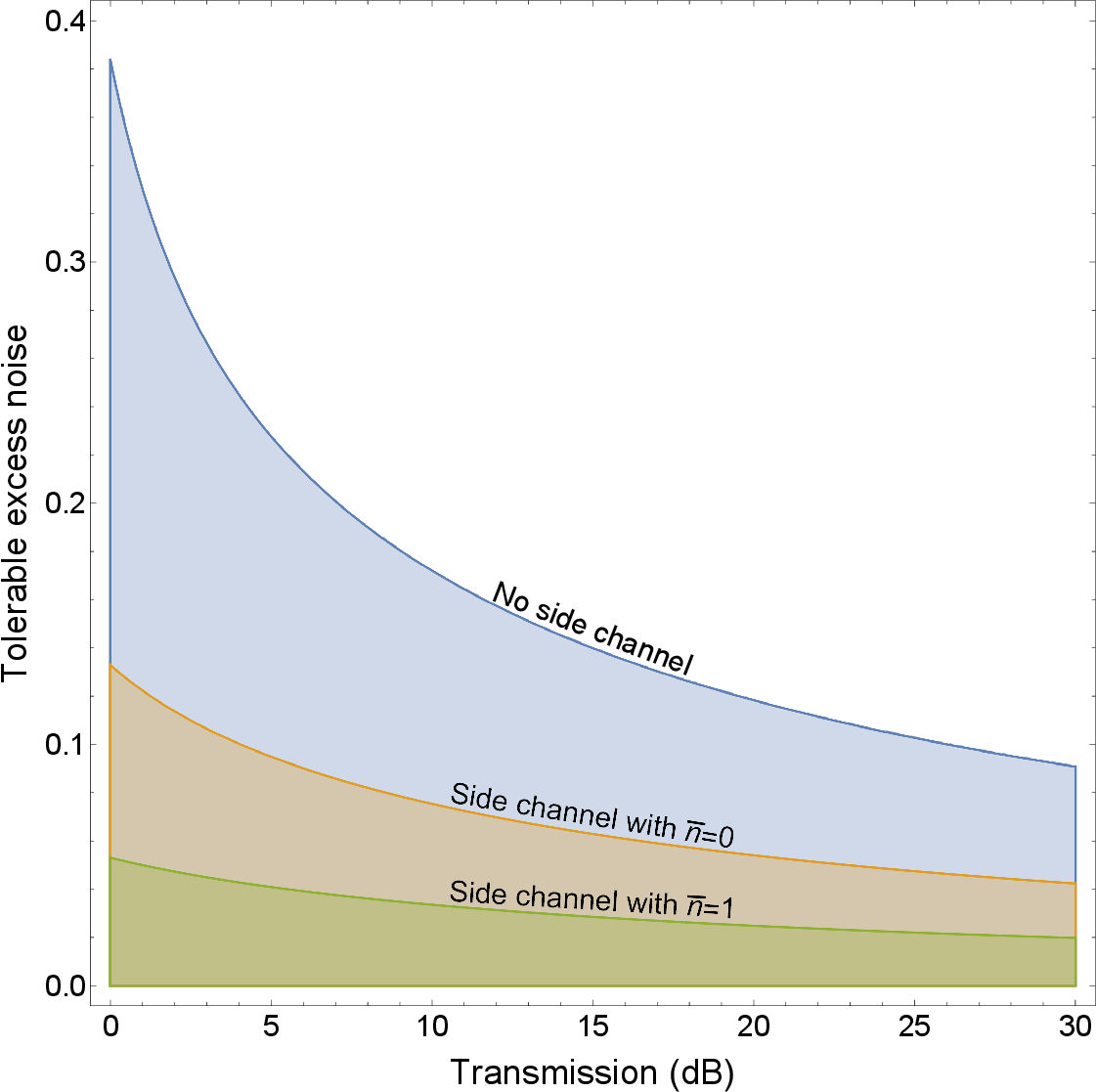}\caption{Security
thresholds in terms of maximally-tolerable excess noise versus channel
transmission (in decibels). The shaded regions are the regions in which secret
key distribution is possible for a given side-channel. The boundaries of the
regions show the values of the excess noise at which secret key distribution
becomes impossible for a given transmission and side-channel. Adding the
leakage mode side-channel significantly decreases the tolerable excess noise
for a given transmission, and increasing the average photon number $\bar{n}$
of the side-channel further decreases it.}%
\label{fig:sec_thresh}%
\end{figure}

We calculate the mutual information between Alice and Bob, $I(\alpha:\beta)$,
as the reduction in (classical) entropy of $\beta$ when conditioned with
$\alpha$. The asymptotic limit of this mutual information is equal to
\begin{align}
I(\alpha:\beta)^{\infty} &  =H(V_{\beta}+1)-H(V_{\beta}|\alpha+1)\\
&  =\log_{2}{\frac{\eta^{\prime}{\mu^{\prime}}}{\eta^{\prime}\epsilon^{\prime
}+2}},
\end{align}
where $H$ is the Shannon entropy~\cite{Cover} and $V_{\beta}$ ($V_{\beta
}|\alpha$) is the variance of Bob's outcome\ $\beta$ (conditional outcome
$\beta|\alpha$). We then calculate the Holevo bound between Eve and Bob in the
asymptotic limit. We find:
\begin{equation}
I(E:\beta)^{\infty}=\log_{2}{\frac{e~{v_{AB,2}^{\infty}}}{2}}+S_{\text{const}%
},
\end{equation}
where
\begin{equation}
S_{\text{const}}=g(v_{AB,1}^{\infty})-g(v_{A|\beta}^{\infty})
\end{equation}
is the entropy contribution that does not scale with $\mu$. The first term of this expression comes from the asymptotic form of $g(v_{AB,2}^{\infty})$, as per Eq.~(\ref{asym}). The asymptotic
secret key rate is given by the difference
\begin{align}
K^{\infty}(\bar{n},\eta,\epsilon) &  =I(\alpha:\beta)^{\infty}-I(E:\beta
)^{\infty}\\
&  =\log_{2}{\frac{2\eta^{\prime}}{e(1-\eta^{\prime})(\eta^{\prime}%
\epsilon^{\prime}+2)}}-S_{\text{const}}.
\end{align}
The extra information gained by Eve due to the side-channel is the difference between the key rate with the side-channel and the key rate without. In general, the asymptotic key rate decreases as the effective transmission
decreases (either due to an increase in the average photon number of the
side-channel mode or due to increased line loss) and as the channel noise
increases. This is shown in the plots in Figs.~\ref{fig:lossy} and
\ref{fig:sec_thresh}.

The asymptotic secret key rate $K^{\infty}$ takes a particularly simple form
if the channel does not add any noise (a pure-loss channel). In fact, it
becomes
\begin{align}
K_{\text{lossy}}^{\infty}  &  =-\frac{\log_{2}{(1-\eta^{\prime})}}%
{\eta^{\prime}}-\log_{2}e\\
&  =\frac{2(\bar{n}+1)}{\eta}\log_{2}\left[  {1-\frac{\eta}{2(\bar{n}+1)}%
}\right]  -\log_{2}e. \label{kkk}%
\end{align}
The rate $K_{\text{lossy}}^{\infty}$ is always positive and plotted in
Fig.~\ref{fig:lossy} for various mean photon numbers $\bar{n}$, where it is
also compared with the ultimate point-to-point rate or PLOB bound $-\log
_{2}(1-\eta)$~\cite{PLOB}. Each time $\bar{n}+1$ doubles (e.g. when $\bar{n}$
goes from 0 to 1, from 1 to 3 or from 3 to 7), the key rate $K_{\text{lossy}%
}^{\infty}$ decreases by approximately 3~dB.

In the low transmission regime (i.e., long distances), it is known that the
PLOB bound becomes roughly linear in $\eta$, and is approximately equal to
$\eta/\ln{2\simeq}1.44\eta$ bits per transmission. It is also known that,
without side-channels, the coherent state protocol has a long-distance ideal
rate of about $\eta/(2\ln{2}){\simeq}0.72\eta$ bits per transmission, which is
half the PLOB bound. The linearity also holds when we include the side
channels. In fact, for low $\eta$, we find that the key rate of Eq.~(\ref{kkk}%
) becomes
\begin{equation}
K_{\text{lossy}}^{\infty}{\simeq\frac{\eta}{4(\bar{n}+1)\ln{2}}\simeq}%
\frac{0.36}{\bar{n}+1}\eta~.
\end{equation}
Note that with the leakage mode ($\bar{n}=0$), this rate is half that of the
coherent state protocol without side-channels. This rate keeps halving each
time $(\bar{n}+1)$ doubles; this can also be seen in the constant decrease in
intercept between each of the plots in Fig.~\ref{fig:lossy}.



We then calculate the threshold excess noise, $\epsilon_{\text{max}}$, for a
given channel transmission, $\eta$, and side-channel parameter, $\bar{n}$.
This is the value of the excess noise up to which secret key distribution is
possible. The threshold condition $\epsilon_{\text{max}}=\epsilon(\eta,\bar
{n})$ is given by solving $K^{\infty}(k,\eta,\epsilon)=0$. In
Fig.~\ref{fig:sec_thresh}, we show the security threshold of the coherent
state protocol~\cite{hetPROT} without side-channels and, then, in two cases
with side-channel modes ($\bar{n}=0$ and $1$). The shaded regions show the
regions in which secret key distribution is possible for a given side-channel.

The leakage mode case ($\bar{n}=0$) has a significantly lower security
threshold than the case with no side-channel, and increasing the average
photon number further decreases the threshold, for fixed transmission. For
instance, for channel transmission of $20$~dB, the presence of leakage
($\bar{n}=0$) decreases the tolerable excess noise by $\simeq0.06$ (from about
$0.12$). For active hacking with $\bar{n}=1$ photon, we have a further
decrease of $\simeq0.03$. In other words, a side-channel with $\bar{n}=1$
gives a $\simeq75\%$ decrease in tolerable excess noise at this distance. If
$\bar{n}$ is increased, the attack becomes even more powerful. It is then
important for Alice to be able to accurately measure $\bar{n}$, by
characterising her devices as accurately as possible.

\subsection{Generalisation of the side-channel}

\begin{figure}[ptb]
\vspace{0.15cm} \centering
\includegraphics[width=1\linewidth]{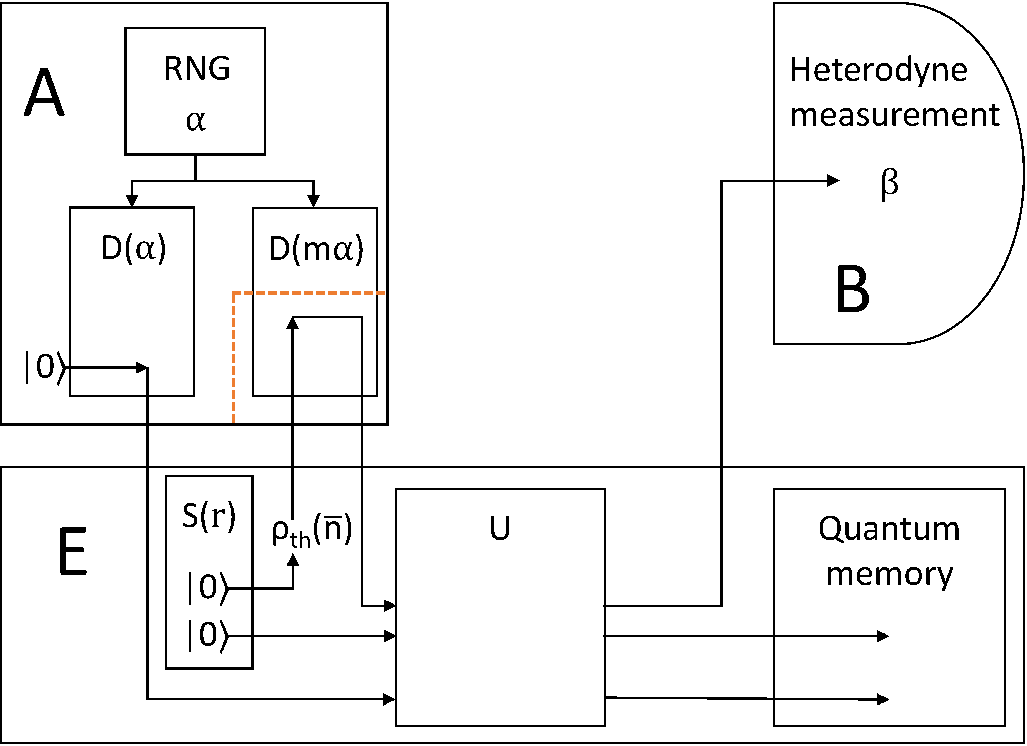}\caption{This is an extension of
the original setup (Fig.~\ref{fig:init_setup}), in which both the average
number of photons entering Alice's device, $\bar{n}$, and the modulation
amplitude of the side-channel mode, $m$, are monitored. Unlike in the original
case, $m$ does not have to equal 1, and can take any real value. The dashed
red line marks the part of Alice's device that is accessible to Eve. The key
rate for this setup can be calculated similarly to the key rate for the
original setup; the only difference is in the expression for the $k$
parameter, which affects the \textquotedblleft effective loss", the
\textquotedblleft effective excess noise" and \textquotedblleft effective
modulation amplitude". See text for more explanation.}%
\label{fig:extend_setup}%
\end{figure}

We can also consider a simple extension, in which Eve's side-channel mode is
modulated by $m\alpha$, whilst Alice's signal state is modulated by $\alpha$.
$m$ is a multiplicative factor on the displacement of the Trojan state; $m=1$ gives the case that has already been considered.
This setup is shown in Fig.~\ref{fig:extend_setup}. Without loss of
generality, we assume that $m>0$, since Eve can always apply a phase shift of
$\pi$ to her modes. Similarly to the original $m=1$ case, we can show that
this attack is equivalent to a standard attack against the main channel but
with an \textquotedblleft effective modulation amplitude", an
\textquotedblleft effective excess noise" and an \textquotedblleft effective
loss". As we show in the appendix, the original and effective parameters are
related by the same Eqs.~(\ref{k1}) and~(\ref{k2}), but where $k$ becomes the
following function of both $\bar{n}$ and $m$~\cite{NotaKAPPA}%
\begin{equation}
k(\bar{n},m)=\sqrt{m^{2}(2\bar{n}+1)+1}.\label{gen_k}
\end{equation}
By monitoring both $\bar{n}$ and $m$, Alice can therefore fully quantify the
effect of any single mode side-channel of this type. Alice can find $\bar{n}$
by monitoring the average photon number entering her device. There are a
number of ways in which she could find $m$. For instance, she could monitor
the total average outgoing photon number of her device across all
modes.

\section{Conclusions and further discussions}

In this work we have considered the effects of hacking Alice's box in one-way
CV\ QKD, namely the coherent state protocol of Ref.~\cite{hetPROT}, which is
hacked while being implemented over a thermal-loss quantum communication
channel. We have assumed that a Trojan horse side-channel mode is introduced
in Alice's device and is modulated in the same way as the signal state. Under
this condition, we have found how quickly the key rate of the original
protocol is deteriorated by increasing the mean number of photons $\bar{n}$
inserted in the device. Even the presence of a leakage mode ($\bar{n}=0$) is
able to halve the rate. Then, each time the value of $(\bar{n}+1)$ doubles,
the long-distance key rate is further halved.

Then we have also considered a direct generalisation of the basic side-channel
attack where the Trojan horse mode is modulated at a different amplitude
($m\alpha$) than the signal state. If this modulation is inefficient ($m<1$),
then the attack is weaker than the basic one. However, if $m>1$, then the
attack becomes more deleterious. In order to deal with this situation, Alice
should be able to estimate not only the mean number of extra photons $\bar{n}$
entering the device, but also the mean number of extra photons leaving the
device, so that she can also evaluate $m$. Therefore, it seems that quantum
metrological tools~\cite{Sam1,Sam2,Paris,Giova,ReviewNEW,nostro,MetroREVIEW}
are necessary inside Alice's box, unless Eve's hacking is mitigated by other
means which suitably modify the original setup and protocol.

If the modulator can be surrounded by a passive attenuator, such that any photons not entering or leaving via the main channel are highly attenuated, the effects of any Trojan horse photons not entering via the main channel can be greatly mitigated. The attenuator can be modelled as a beamsplitting operation with a vacuum mode. The effects on the information gained by Eve are twofold: the first quadrature of her side-channel mode is scaled down by the attenuation and the correlations between Eve's Trojan horse mode and her idler mode are reduced. The conditional state $\psi_{E_S,E_I}|\alpha$ received by Eve after the side-channel (conditioned on Alice's value), is an attenuated TMSV state. It has the covariance matrix
\begin{equation}
V_{E_S,E_I}|\alpha=\begin{pmatrix}
(2\bar{n}+1)\mathbf{1} &2\sqrt{\bar{n'}(\bar{n}+1)}\mathbb{Z}\\
2\sqrt{\bar{n'}(\bar{n}+1)}\mathbb{Z} &(2\bar{n'}+1)\mathbf{1}
\end{pmatrix},
\end{equation}
where $\bar{n}$ is the photon number of the TMSV state prior to the side channel, $\bar{n'}$ is some positive real number less than $\bar{n}$, $E_S$ is the signal mode, and $E_I$ is the idler mode.

Let $\psi_{E_S E_I P}|\alpha$ be the purification of $\psi_{E_S,E_S}|\alpha$. Eve does not hold the purifying mode $P$, but if she did, it could only help her. Therefore, let us assume she is given it; this is equivalent to saying she is given the other output of the beamsplitter with its quadratures set to 0. Then the modes $E_I P$ together purify $E_S$. Any one-mode thermal state can be purified by a TMSV~\cite{weedbrook_gaussian_2012}. Hence, there exists some unitary acting only on the purifying systems $E_I$ and $P$ that results in a TMSV on the modes $E_S E_I$ with $\bar{n'}$ photons per mode (and a vacuum state on $P$). If the Trojan horse mode is modulated by $m\alpha$, we can say that the first quadrature of this mode after the attenuator is $m'\alpha$, where $m'$ is some positive, real number less than $m$. Then, the key rate is lower-bounded by the key rate calculated before, but with $\bar{n}$ and $m$ replaced by $\bar{n'}$ and $m'$ respectively. More specifically, the expression for $k$ in Eq.~(\ref{gen_k}) becomes
\begin{align}
k(\bar{n'},m')&=\sqrt{m'^{2}(2\bar{n'}+1)+1}\\
&=\sqrt{T m^{2}(2T^2\bar{n'}+1)+1},
\end{align}
where $T$ is the transmission of the attenuator. This expression rapidly approaches unity as $T$ decreases. Here we have assumed that the Trojan horse state passes through the attenuator twice: once prior to modulation and once after modulation. Hence, we have set $m'=\sqrt{T}m$ and $\bar{n'}=T^2\bar{n}$.

The expression for the maximum secret key rate in this case is a lower bound: giving Eve access to the purification mode $P$ cannot decrease the Holevo bound on her mutual information, but it is not immediately obvious whether it increases it. It is therefore not obvious whether this lower bound is tight, or whether the power of the side-channel would be even further reduced by the attenuation; this is a question that is open to further study.

Without upper-bounding the incoming photon number, the addition of an attenuator does not provide provable security by itself, since we do not know the initial values of $m$ and $\bar{n}$, hence quantum metrological tools are still required. It may be possible to find an upper-bound on the incoming photon number for a given device, using physical considerations, such as the point at which damage to the modulator from the incoming photons would become obvious to Alice.

In order to limit the effects of a Trojan horse mode introduced via the main channel, the passive architecture to limit Trojan horse attacks in the DV case, introduced in \cite{lucamarini_practical-security_2015}, could be implemented in the CV case. The incoming photon number is bounded using the LIDT of the optical fibre constituting the main quantum channel; the photon number threshold is dependent on the frequency of the incoming photons, since lower frequency photons are less energetic, however the frequency is bounded from below by an optical fibre loop and a filtering block, which select for frequencies. There is then an attenuator, which greatly reduces the photon number of any incoming state. In this case, the attenuation will not decrease the magnitude of modulation of the Trojan horse state compared to the signal state, as it does in the case in which the Trojan horse photons do not enter via the main channel. This is because the signal state will be attenuated in the same way as the Trojan horse state. The incoming photons sent by Eve would still be attenuated, leading to damping of the off-diagonal elements of $V_{E1,E2}|\alpha$. By suitably choosing the attenuation, bearing in mind the maximum photon number of Eve's Trojan horse state, Alice could decrease the correlations between Eve's Trojan horse mode and her idler mode to an arbitrary degree, and hence could effectively reduce Eve's side-channel to a leakage mode ($\bar{n}=0$).

Bounding the incoming photon number using the LIDT raises another issue, since we have assumed that modulation of the signal mode is unbounded, and hence can be taken to infinity. Since the photon number of the signal state will also be limited by the LIDT, this is not entirely true. However, if the LIDT is sufficiently high, this should not greatly affect the secret key rate. Increasing the LIDT does not increase the photon number of Eve's outgoing side-channel mode as long as the attenuation is raised accordingly.

One further possible problem could occur if the attenuator itself is not properly characterised. If it re-radiates absorbed photons, or scatters light in such a way that it is accessible to Eve, the attenuator itself may provide a leakage side-channel. Alternatively, it may have a much higher transmission at certain frequencies, allowing Eve to send Trojan horse photons through without much attenuation.

\smallskip

\acknowledgments This work was made possible via the EPSRC Quantum
Communications Hub (Grant No. EP/M013472/1). The authors would like to thank Pieter Kok, Scott E. Vinay, Panagiotis Papanastasiou, Romain All\'{e}aume, Marco Lucamarini, Rupesh Kumar, and Vladyslav Usenko, for comments and feedback.

\appendix

\section{Calculation of the k-value for any m-value}

The steps to study the setup in Fig.~\ref{fig:extend_setup} are very similar
to those for the $m=1$ case. By using a beamsplitter on modes 1 and 2 followed
by two-mode squeezers on modes 2 and 3 and then on modes 1 and 3, we can show
that the setup is equivalent to one in which the signal state is modulated by
$k_{1}\alpha$ and a single pure side-channel mode is modulated by
$k_{2}\mathbb{Z}\alpha$ (as in Fig.~\ref{fig:equiv1}, but with different
values for $k_{1}$ and $k_{2}$). We then again use the fact that this gives
the same key rate as a setup in which the side-channel mode is modulated by
$k_{2}\alpha$ instead of by $k_{2}\mathbb{Z}\alpha$, and hence that it gives
the same key rate as one in which the signal state is modulated by
$k=\sqrt{k_{1}^{2}+k_{2}^{2}}$, with a beamsplitter in the main channel.

We label the initial covariance matrix of the total state as $V_{0}^{m\neq1}$,
the initial covariance matrix for fixed $\alpha$ as $V_{0}^{m\neq1}|\alpha$
and the initial quadratures for fixed $\alpha$ as $X_{0}^{m\neq1}|\alpha$, and
then use the subscripts 1, 2 and 3 to denote these objects after the
beamsplitter, the first two-mode squeezer and the second two-mode squeezer
respectively. The optical circuit is the same as in Fig.~\ref{fig:circuit};
only the parameters of the optical components are changed for the $m\neq1$ case.

The first and second moments of the initial state are
\begin{align}
X_{0}^{m\neq1}|\alpha &  =%
\begin{pmatrix}
\alpha\\
m\alpha\\
0
\end{pmatrix}
,\\
V_{0}^{m\neq1}|\alpha &  =%
\begin{pmatrix}
\mathbf{1} & \mathbf{0} & \mathbf{0}\\
\mathbf{0} & \cosh{2r}\mathbf{1} & \sinh{2r}\mathbb{Z}\\
\mathbf{0} & \sinh{2r}\mathbb{Z} & \cosh{2r}\mathbf{1}%
\end{pmatrix}
,\\
V_{0}^{m\neq1}  &  =%
\begin{pmatrix}
(\mu+1)\mathbf{1} & m\mu\mathbf{1} & \mathbf{0}\\
m\mu\mathbf{1} & (m^{2}\mu+\cosh{2r})\mathbf{1} & \sinh{2r}\mathbb{Z}\\
\mathbf{0} & \sinh{2r}\mathbb{Z} & \cosh{2r}\mathbf{1}%
\end{pmatrix}
.
\end{align}

The first optical component is a beamsplitter that sets the quadratures of
modes 2 and 3 to 0 (moves the entire displacement onto mode 1). This
beamsplitter has angle
\begin{equation}
\theta_{1}^{m\neq1}=\arccos{\frac{1}{\sqrt{m^{2}+1}}},
\end{equation}
and changes the first and second moments of the state to
\begin{align}
X_{1}^{m\neq1}|\alpha &  =%
\begin{pmatrix}
\sqrt{m^{2}+1}\alpha\\
0\\
0
\end{pmatrix}
,\\
V_{1}^{m\neq1}|\alpha &  =%
\begin{pmatrix}
\frac{m^{2}\cosh{2r}+1}{m^{2}+1}\mathbf{1} & \frac{2m\sinh^{2}{r}}{m^{2}%
+1}\mathbf{1} & my^{(1)}\mathbb{Z}\\
\frac{2m\sinh^{2}{r}}{m^{2}+1}\mathbf{1} & \frac{m^{2}+\cosh{2r}}{m^{2}%
+1}\mathbf{1} & y^{(1)}\mathbb{Z}\\
my^{(1)}\mathbb{Z} & y^{(1)}\mathbb{Z} & \cosh{2r}\mathbf{1}%
\end{pmatrix}
,\\
V_{1}^{m\neq1} &  =V_{1}^{m\neq1}|\alpha\oplus(m^{2}+1)\mu%
\begin{pmatrix}
\mathbf{1} &  & \\
& \mathbf{0} & \\
&  & \mathbf{0}%
\end{pmatrix}
,
\end{align}
where $y^{(1)}=(\sinh{2r)/}\sqrt{m^{2}+1}$.

The next component purifies the second mode, reducing the state to a bipartite
state. It acts on the second and third modes and has squeezing parameter
$r_{2}^{m\neq1}=-\arcsinh{\frac{\sqrt{2}\sinh{r}}{\sqrt{m^2\cosh{2r}+m^2+2}}}$%
. The first and second moments become
\begin{align}
X_{2}^{m\neq1}|\alpha &  =%
\begin{pmatrix}
\sqrt{m^{2}+1}\alpha\\
0\\
0
\end{pmatrix}
,\\
V_{2}^{m\neq1}|\alpha &  =%
\begin{pmatrix}
\frac{m^{2}\cosh{2r}+1}{m^{2}+1}\mathbf{1} & \mathbf{0} & y^{(2)}\mathbb{Z}\\
\mathbf{0} & \mathbf{1} & \mathbf{0}\\
y^{(2)}\mathbb{Z} & \mathbf{0} & \frac{m^{2}\cosh{2r}+1}{m^{2}+1}\mathbf{1}%
\end{pmatrix}
,\\
V_{2}^{m\neq1} &  =V_{2}^{m\neq1}|\alpha\oplus(m^{2}+1)\mu%
\begin{pmatrix}
\mathbf{1} &  & \\
& \mathbf{0} & \\
&  & \mathbf{0}%
\end{pmatrix}
,
\end{align}
where
\begin{equation}
y^{(2)}=\frac{\sqrt{2}m\sinh{r}\sqrt{m^{2}\cosh{2r}+m^{2}+2}}{m^{2}+1}.
\end{equation}

The final component unsqueezes the remaining two modes, such that the state
for fixed $\alpha$ is a vacuum state. The squeezing parameter is $r_{3}%
^{m\neq1}=-\arcsinh{\frac{m\sinh{r}}{\sqrt{m^2+1}}}$. The first and second
moments become
\begin{align}
X_{3}^{m\neq1}|\alpha &  =%
\begin{pmatrix}
\frac{\sqrt{m^{2}\cosh{2r}+m^{2}+2}}{\sqrt{2}}\alpha\\
-m\sinh{r}\mathbb{Z}\alpha\\
0
\end{pmatrix}
=%
\begin{pmatrix}
k_{1}^{m\neq1}\alpha\\
k_{2}^{m\neq1}\mathbb{Z}\alpha\\
0
\end{pmatrix}
,\\
V_{3}^{m\neq1}|\alpha &  =%
\begin{pmatrix}
\mathbf{1} & \mathbf{0}\\
\mathbf{0} & \mathbf{1}%
\end{pmatrix}
,~~V_{3}^{m\neq1}=%
\begin{pmatrix}
x_{+}\mathbf{1} & y^{(3)}\mathbb{Z}\\
y^{(3)}\mathbb{Z} & x_{-}\mathbf{1}%
\end{pmatrix}
,
\end{align}
where
\begin{align}
x_{\pm} &  =\frac{1}{2}(m^{2}\mu\cosh{2r}\pm m^{2}\mu+2),\\
y^{(3)} &  =-\frac{m\mu\sinh{r}\sqrt{m^{2}\cosh{2r}+m^{2}+2}}{\sqrt{2}}.
\end{align}

Since we have shown that there is an optical circuit that reversibly converts
the initial state of the setup in Fig.~\ref{fig:extend_setup} to the initial
state of the setup in Fig.~\ref{fig:equiv1}, the two setups must have the same
secret key rate for the same thermal noise. As shown in the main text, this
also means that the setup in Fig.~\ref{fig:extend_setup} has the same secret
key rate as the side-channel-free setup with an \textquotedblleft effective
modulation\textquotedblright\ of $\mu^{eff}=k^{2}\mu$, an \textquotedblleft
effective channel loss\textquotedblright\ of $\eta^{eff}=\frac{\eta}{k^{2}}$
and an \textquotedblleft effective excess noise\textquotedblright\ of
$\epsilon^{eff}=k^{2}\epsilon$, where
\begin{align}
k  &  =\sqrt{k_{1}^{2}+k_{2}^{2}}\\
&  =\sqrt{\frac{1}{2}(m^{2}\cosh{2r}+m^{2}+2)+m^{2}\sinh^{2}{r}}\\
&  =\sqrt{m^{2}(2\bar{n}+1)+1}.
\end{align}
This is the result given in the main text.


\begin{thebibliography}{99}                                                                                               %


\bibitem {Watrous}J. Watrous, \textit{The theory of quantum information}
(Cambridge University Press, Cambridge, 2018).

\bibitem {Hayashi}M. Hayashi, \textit{Quantum Information Theory: Mathematical
Foundation} (Springer-Verlag Berlin Heidelberg, 2017).

\bibitem {HolevoBOOK}A. Holevo,\textit{ Quantum Systems, Channels,
Information: A Mathematical Introduction} (De Gruyter, Berlin-Boston, 2012).

\bibitem {NC00}M. A. Nielsen and I. L. Chuang, \textit{Quantum Computation and
Quantum Information} (Cambridge University Press, Cambridge, 2000).

\bibitem {rev1}N. Gisin, G. Ribordy, W. Tittel, and H. Zbinden, Rev. Mod.
Phys. \textbf{74}, 145-196 (2002).

\bibitem {rev2}V. Scarani, H.~Bechmann-Pasquinucci, N.~J.~Cerf, M.~Dusek,
N.~Lutkenhaus, M.~Peev, Rev. Mod. Phys. \textbf{81}, 1301 (2008).

\bibitem {rev3}E. Diamanti and A. Leverrier, Entropy \textbf{17}, 6072-6092 (2015).

\bibitem {pironio_device-independent_2009}S. Pironio, A. Acin, N. Brunner, N.
Gisin, S. Massar, and V. Scarani, New J. Phys. \textbf{11}, 045021 (2009).

\bibitem {BB84}C. H. Bennett, and G. Brassard, Proc. IEEE International Conf.
on Computers, Systems, and Signal Processing, Bangalore, pp. 175--179 (1984).

\bibitem {decoy}W.-Y. Hwang, Phys. Rev. Lett. \textbf{91}, 057901(2003).

\bibitem {lo_decoy_2005}H.-K. Lo, X. Ma, and K. Chen, Phys. Rev. Lett.
\textbf{94}, 230504 (2005).

\bibitem {scarani_black_2009}V. Scarani and C. Kurtsiefer, Theor. Comput. Sci.
\textbf{560}, 27 (2014).

\bibitem {lydersen_hacking_2010}L. Lydersen, C. Wiechers, C. Wittmann, D.
Elser, J. Skaar, and V. Makarov, Nat. Photon. \textbf{4}, 686 (2010).

\bibitem {zhao_quantum_2008}Y. Zhao, C.-H. F. Fung, B. Qi, C. Chen, and H.-K.
Lo, Phys. Rev. A \textbf{78}, 042333 (2008).

\bibitem {jain_trojan-horse_2014}N. Jain, E. Anisimova, I. Khan, V. Makarov,
C. Marquardt, and G. Leuchs, New J. Phys. \textbf{16}, 123030 (2014).

\bibitem{haseler_testing_2008}H.~H\"aseler, T.~Moroder, and N.~L\"utkenhaus, Phys. Rev. A \textbf{77}, 032303 (2008).

\bibitem{huang_quantum_2013}J.-Z. Huang, C.~Weedbrook, Z.-Q. Yin, S.~Wang, H.-W. Li, W.~Chen, G.-C. Guo, and Z.-F. Han, Phys. Rev. A \textbf{87}, 062329 (2013).

\bibitem{ma_wavelength_2013}X.-C. Ma, S.-H. Sun, M.-S. Jiang, and L.-M. Liang, Phys. Rev. A \textbf{87}, 052309 (2013).

\bibitem{jouguet_preventing_2013}P.~Jouguet, S.~Kunz-Jacques, and E.~Diamanti, Phys. Rev. A \textbf{87}, 062313 (2013).

\bibitem{huang_quantum_2014}J.-Z. Huang, S.~Kunz-Jacques, P.~Jouguet, C.~Weedbrook, Z.-Q. Yin, S.~Wang, W.~Chen, G.-C. Guo, and Z.-F. Han, Phys. Rev. A \textbf{89}, 032304 (2014).

\bibitem{qin_quantum_2016}H.~Qin, R.~Kumar, and R.~All\'{e}aume, Phys. Rev. A \textbf{94}, 012325 (2016).

\bibitem {qin_homodyne_2018}H. Qin, R. Kumar, V. Makarov, and R. All\'{e}aume, Phys. Rev. A \textbf{98}, 012312 (2018).

\bibitem {Ekert}A. K. Ekert, Phys. Rev. Lett. \textbf{67}, 661--663 (1991).

\bibitem {samMDI2012}S. L. Braunstein, and S. Pirandola, Phys. Rev. Lett.
\textbf{108}, 130502 (2012).

\bibitem {lo_measurement-device-independent_2012}H.-K. Lo, M. Curty, and B.
Qi, Phys. Rev. Lett. \textbf{108}, 130503 (2012).

\bibitem {pirandola_high-rate_2015}S. Pirandola, C. Ottaviani, G. Spedalieri,
C. Weedbrook, S. L. Braunstein, S. Lloyd, T. Gehring, C. S. Jacobsen, and U.
L. Andersen, Nat. Photon. \textbf{9}, 397 (2015).

\bibitem {vakhitov_large-pulse_2001}A. Vakhitov, V. Makarov, and D. R. Hjelme, J. Mod. Opt.
\textbf{48}, 2023 (2001).

\bibitem {vinay_burning_2018}S. E. Vinay and P. Kok, Phys. Rev. A \textbf{97},
042335 (2018).

\bibitem {gisin_trojan_2006}N. Gisin, S. Fasel, B. Kraus, H. Zbinden, and G. Ribordy, Phys. Rev. A
\textbf{73}, 022320 (2006).

\bibitem {lucamarini_practical-security_2015}M. Lucamarini, I. Choi, M. B. Ward, J. F. Dynes, Z. L. Yuan, and A. J. Shields, Phys. Rev. X
\textbf{5}, 031030 (2015).

\bibitem {tamaki_decoy-state_2016}K. Tamaki, M. Curty, and M. Lucamarini, New J. Phys.
\textbf{18}, 065008 (2016).

\bibitem {hetPROT}C. Weedbrook, A. M. Lance, W. P. Bowen, T. Symul, T. C.
Ralph, and P. K. Lam, Phys. Rev. Lett. \textbf{93}, 170504 (2004).

\bibitem {jain_risk-analysis_2014}N. Jain, B. Stiller, I. Khan, V. Makarov, C. Marquadt, and G. Leuchs, IEEE J Sel Top Quantum Electron.
\textbf{21}, 168 (2014).

\bibitem {weedbrook_gaussian_2012}C. Weedbrook, S. Pirandola, R.
Garcia-Patron, N. J. Cerf, T. C. Ralph, J. H. Shapiro, and S. Lloyd, Rev. Mod.
Phys. \textbf{84}, 621 (2012).

\bibitem {grosshans_virtual_2003}F. Grosshans, N.~J. Cerf, J.~Wenger,
R.~Tualle-Brouri, and Ph~Grangier, Quantum Info. Comput. \textbf{3}, 535 (2003).

\bibitem {usenko_trusted_2016}V. C. Usenko and R. Filip, Entropy \textbf{18},
20 (2016).

\bibitem {derkach_continuous-variable_2017}I. Derkach, V. C. Usenko, and R.
Filip, Phys. Rev. A \textbf{96}, 062309 (2017).

\bibitem {derkach_preventing_2016}I. Derkach, V. C. Usenko, and R.
Filip, Phys. Rev. A \textbf{93}, 032309 (2016).

\bibitem {pirandola_characterization_2008}S. Pirandola, S. L. Braunstein, and
S. Lloyd, Phys. Rev. Lett. \textbf{101}, 200504 (2008).

\bibitem {grosshans_collective_2005}F. Grosshans, Phys. Rev. Lett.
\textbf{94}, 020504 (2005).

\bibitem {PLOB}S. Pirandola, R. Laurenza, C. Ottaviani, and L. Banchi, \ Nat.
Commun. \textbf{8}, 15043 (2017). See also arXiv:1510.08863.

\bibitem {TQC}S. Pirandola, S. L. Braunstein, R. Laurenza, C. Ottaviani, T. P.
W. Cope, G. Spedalieri, and L. Banchi, Quantum Sci. Technol. \textbf{3},
035009 (2018).

\bibitem {Cover}T. M. Cover, and J. A. Thomas, \textit{Elements of Information
Theory} (2nd edition, Wiley, 2006).

\bibitem {NotaKAPPA}Note that if $m=1$, this reduces to the previous case.
Note also that if $m=0$, we do not have a side-channel and so $k=1$, hence the
\textquotedblleft effective loss\textquotedblright\ is equal to the observed
loss, as we would expect.

\bibitem {Sam1}S. L. Braunstein and C. M. Caves, Phys. Rev. Lett. \textbf{72},
3439 (1994).

\bibitem {Sam2}S. L. Braunstein, C. M. Caves, and G. J. Milburn, Ann. Phys.
\textbf{247}, 135-173 (1996).

\bibitem {Paris}M. G. A. Paris, Int. J. Quant. Inf. \textbf{7}, 125-137 (2009).

\bibitem {Giova}V. Giovannetti, S. Lloyd, and L. Maccone, Nature Photon.
\textbf{5}, 222 (2011).

\bibitem {ReviewNEW}D. Braun, G. Adesso, F. Benatti, R. Floreanini, U. Marzolino, M. W. Mitchell, and S. Pirandola, Rev. Mod. Phys. \textbf{90}, 035006 (2018).

\bibitem {nostro}S. Pirandola, and C. Lupo, Phys. Rev. Lett. \textbf{118},
100502 (2017); \textit{ibidem} \textbf{119}, 129901 (2017).

\bibitem {MetroREVIEW}R. Laurenza, C. Lupo, G. Spedalieri, S. L. Braunstein,
and S. Pirandola, Quantum Meas. Quantum Metrol. \textbf{5}, 1--12 (2018).
\end{thebibliography}
\end{document}